\documentclass[journal]{IEEEtran}
\ifCLASSINFOpdf
\else
\fi
\usepackage{amsmath} 
\usepackage{amssymb} 
\usepackage{graphicx}
\usepackage{comment}
\usepackage{cite}
\usepackage{float}
\usepackage{xcolor} 
\usepackage[ruled,vlined,linesnumbered]{algorithm2e}
\DontPrintSemicolon
\SetKwInOut{KwInput}{Input}
\SetKwInOut{KwOutput}{Output}

\begin{document}

\title{Video TokenCom: Textual Intent-guided Multi-Rate Video Token Communications with UEP-based Adaptive Source–Channel Coding}

\author{Jingxuan~Men,~\IEEEmembership{Student~Member,~IEEE,}
        Mahdi~Boloursaz~Mashhadi,~\IEEEmembership{Senior~Member,~IEEE,}
        Ning~Wang,~\IEEEmembership{Senior~Member,~IEEE,}
        Yi~Ma,~\IEEEmembership{Senior~Member,~IEEE,}
        Mike~Nilsson,
        and~Rahim~Tafazolli,~\IEEEmembership{Fellow,~IEEE}
\thanks{%
J. Men, M. Boloursaz Mashhadi, Y. Ma, and R. Tafazolli are with 5GIC \& 6GIC,
Institute for Communication Systems (ICS), University of Surrey, U.K. (e-mail: {j.men, m.boloursazmashhadi, y.ma, r.tafazolli}@surrey.ac.uk).\\
\hspace*{1em}N. Wang is with the Smart Internet Lab, University of Bristol, U.K. (e-mail: n.wang@bristol.ac.uk).\\
\hspace*{1em}M. Nilsson is with the British Telecom Research Lab, BT Group plc, U.K. (e-mail: mike.nilsson@bt.com).}}

\maketitle

\begin{abstract}
Token Communication (TokenCom) is a new paradigm, motivated by the recent success of Large AI Models (LAMs) and Multimodal Large Language Models (MLLMs), where tokens serve as unified units of communication and computation, enabling efficient semantic- and goal-oriented information exchange in future wireless networks. In this paper, we propose a novel Video TokenCom framework for textual intent-guided multi-rate video communication with Unequal Error Protection (UEP)-based source-channel coding adaptation. The proposed framework integrates user-intended textual descriptions with discrete video tokenization and unequal error protection to enhance semantic fidelity under restrictive bandwidth constraints. First, discrete video tokens are extracted through a pretrained video tokenizer, while text-conditioned vision-language modeling and optical-flow propagation are jointly used to identify tokens that correspond to user-intended semantics across space and time. Next, we introduce a semantic-aware multi-rate bit-allocation strategy, in which tokens highly related to the user intent are encoded using full codebook precision, whereas non-intended tokens are represented through reduced codebook precision differential encoding, enabling rate savings while preserving semantic quality. Finally, a source and channel coding adaptation scheme is developed to adapt bit allocation and channel coding to varying resources and link conditions. Experiments on various video datasets demonstrate that the proposed framework outperforms both conventional and semantic communication baselines, in perceptual and semantic quality on a wide SNR range.
\end{abstract}


\IEEEpeerreviewmaketitle
\bstctlcite{IEEEexample:BSTcontrol}

\section{Introduction}
\IEEEPARstart{T}{okens} have become the fundamental units of representation and processing in modern generative foundation models, for perception, reasoning, and content generation across modalities~\cite{qiao2025tokencom, qiao2025todma}. Discrete token representations enable high-dimensional signals to be compressed into compact and semantically meaningful sequences, as demonstrated by vector-quantized and discrete latent modeling approaches in vision and multimodal learning~\cite{qiao2025tokencom, qiao2025todma}. With the rapid development of Multimodel Large Language Models (MLLMs), learned tokenization mechanisms further support semantic alignment across text and vision, preserving both perceptual details and high-level semantic concepts~\cite{brown2020gpt3, radford2021clip,huang2023language}. These characteristics suggest a natural interface between token-based representations and wireless communication systems, where discrete, semantically structured information units can be directly mapped onto resource-constrained transmission pipelines. Motivated by this convergence between foundation models and wireless communications, token-based communication has emerged as a new design paradigm that departs from conventional bit-level transmission and instead operates on semantically meaningful token representations. 
The TokenCom framework was first proposed in~\cite{qiao2025tokencom}, aiming to address the lack of generalizability gap in the conventional semantic- and goal-oriented communications \cite{9955312, 9955525, 10872776}. TokenCom establishes a context-aware architecture that leverages transformer-based foundation models to extract, transform, and transmit multimodal information directly in the token domain. Subsequent works have advanced token-centric wireless technologies across multiple dimensions, including token-domain multiple access~\cite{11152964, qiao2025todma}, semantic packet aggregation via genetic or lookahead search~\cite{lee2025spa,lee2025lookahead}, adaptive token selection for goal-oriented or transformer-based edge inference~\cite{devoto2024adaptive,devoto2025adaptive}, text-guided token-based image communication~\cite{liu2025textguided}, multimodal semantic communication with selective training tokens~\cite{peng2025selective}, and loss-resilient masked-visual-token modeling~\cite{wang2025resicomp}. TokenCom has further been studied from an information-bottleneck perspective~\cite{wei2025ib} and extended to resource-constrained multiuser networks empowered by multimodal large models~\cite{zhang2025mllm}. Very recently, joint semantic–channel coding and adaptive modulation strategies was investigated for TokenCom with a cross-layer design between token semantics and physical-layer transmission~\cite{jointsemchan2025}. Finally, TokenCom-UEP~\cite{tokcomuep2025}, demonstrating improved robustness by assigning differentiated channel protection to semantically important tokens. Despite these advances, video TokenCom remains less studied.  

Most existing video semantic communication systems operate on continuous feature representations or task-specific latent codes, and do not explicitly exploit discrete video tokens as unified units of compression and communication. Although discrete video tokenization has emerged as an effective representation for high-dimensional video data, existing frameworks do not exploit token-level semantic structure for efficient communications, e.g. optimal allocation of the video source and channel coding rate. In this paper, we address these gaps by proposing a Video Token Communications with multi-rate textual intent-guided source-channel coding adaptation framework that integrates video tokenization, text-conditioned intent-relevance extraction, multi-rate bit allocation strategy, and semantic UEP-based source--channel adaptation. Different from many existing end-to-end learned or deep joint source channel coding (DJSCC)-base semantic communication schemes \cite{SemCom1, SemCom2}, the proposed framework is designed based on source-channel separation thanks to pre-trained discrete tokenization, which facilitates flexibility under different channel/network conditions, and an ITU-T OSI compatible multilayer design enabling scalability. We first decompose the video into Intra-coded frames (I-frames) and Predictive-coded frames (P-frames), and then employ a pre-trained video tokenizer to map video frames into discrete tokens. We then leverage a vision language model to identify user-intended semantic regions of the video across space and time. These semantic masks are mapped onto the token grid, yielding user-intended and non-intended token sets explicitly associated with the textual intent. Next, we design a semantic-aware multi-rate bit allocation scheme in which tokens within user-intended regions are transmitted with full bit-precision, whereas non-intended tokens are differentially-coded at reduced precision relative to a reference frame, thereby saving bitrate while preserving the quality of intent-relevant content. Finally, we formulate a source–channel coding adaptation problem under a fixed resource budget, assigning source bit-precision and modulation–coding configurations separately to intended and non-intended token classes to balance task distortion and transmission delay within a UEP framework. The main contributions of this work are thereby summarized as follows:

\begin{itemize}
    \item We develop an intent-relevance extraction framework based on vision language modeling and optical-flow trajectories on video frames, that leverages user textual intent descriptions to convert video content into discrete token classes explicitly associated with the user-intended and non-intended semantics. 

    \item To achieve semantic-aware source coding, we propose a multi-rate bit-allocation strategy that assigns different source coding bit-precisions to distinct video token classes, i.e. intended vs. non-intended, based on their semantic relevance. The intended token class is encoded using full token codebook precision bits, while the non-intended class is encoded differentially with reduced precision via a compact reduced codebook. This design significantly improves rate efficiency while preserving the visual quality of intent-relevant semantic regions. Simulation results show that the proposed TokenCom framework outperforms both conventional H.265~\cite{hevc, ISO23008-2:2025} and generative diffusion-based VC-DM~\cite{videocompress_bench} video coding baselines at ultra-low bits per pixel (BPP) values.

    \item We formulate a source-channel optimization scheme that explicitly balances semantic distortion and end-to-end transmission delay, under limited resource constraints. Following a UEP approach, the proposed scheme adapts the modulation and coding scheme (MCS) separately for the intended and non-intended token classes, where all tokens in the same class share the same MCS. The optimization selects one configuration per token class to minimize a weighted sum of distortion and delay, subject to a fixed resource budget and reliability constraints. Compared with conventional H.265 video compression and transmission across various SNR values, Video TokenCom achieves consistent gains in peak signal-to-noise ratio (PSNR), structural similarity index measure (SSIM), learned perceptual image patch similarity (LPIPS), Fréchet video distance (FVD), and CLIP-based semantic similarity (CLIP), reducing FVD by nearly 1500 at an SNR of 6 dB.

\end{itemize}
The rest of this paper is organized as follows. Section~II introduces the proposed textual-intent-guided multi-rate video token communication framework and the UEP-based adaptive source–channel coding system model, along with the problem formulation. Sections~III and~IV present the experimental evaluation and ablation studies, respectively. Finally, conclusions are drawn in Section~V.

\section{System Model}
\begin{figure*}[t]
    \centering
    \includegraphics[width=0.9\textwidth]{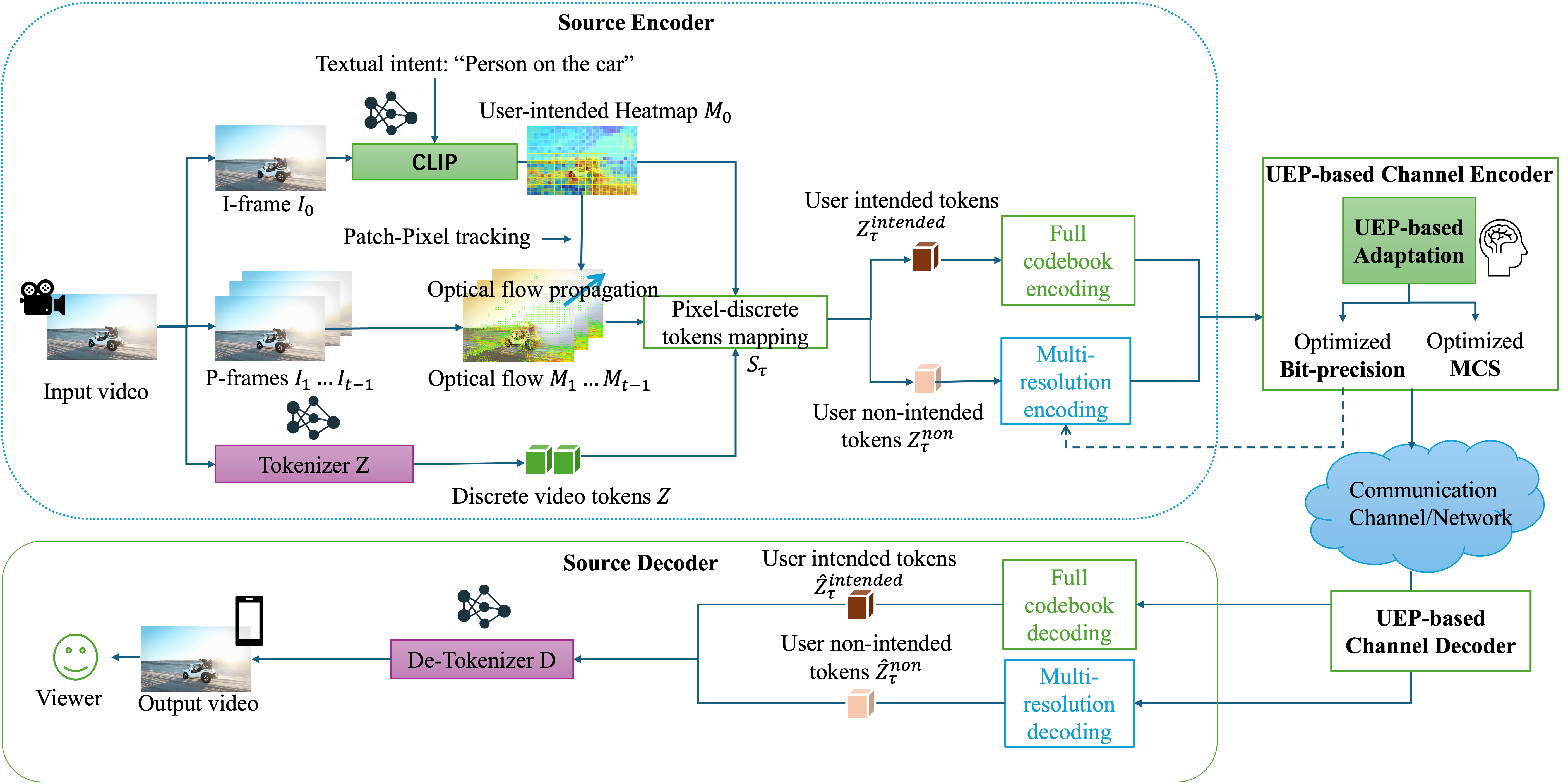} 
    \caption{Overall architecture of the proposed Video Token Communications framework with Multi-rate Textual Intent Source-Channel Coding Adaptation. The pipeline consists of: (1) proposed token-based textual intent-guided source encoder with multimodal user-intended token extraction and semantic-aware multi-rate bit coding; (2) proposed token-based source decoder; and (3) proposed UEP-based source channel coding/decoding adaptation.}
    \label{pic:Architecture} 
\end{figure*}

Fig.~\ref{pic:Architecture} illustrates the overall system architecture, which consists of three main components: (i) a token-based textual intent-guided source encoder with multimodal user-intended token extraction and semantic-aware multi-rate bit coding, (ii) a token-based source decoder, and (iii) a UEP-based source channel coding/decoding adaptation. The proposed architecture offers flexibility to allow different entities to issue the textual intent, either from content producers or consumers. For instance, a content provider can embed textual intent with individual video segments in order to guide adaptive source channel coding when being streamed under various channel conditions. On the other hand, allowing content consumers to express the textual intent  during content consumption sessions may offer higher flexibility in terms of supporting potentially different viewing focuses, but at a small intent feedback overhead.

\subsection{Video Tokenization}
We deploy a video tokenizer to convert the video into discrete token representations. Given an input video $\mathbf{V} \in \mathbb{R}^{T \times H \times W}$ with number of frames (temporal length) $T$, and spatial resolution $H \times W$, we apply a pre-trained video tokenizer $\mathcal{E}(\cdot)$ with a pre-learned discrete token codebook $\mathbf{C} = \{\mathbf{c}_1, \mathbf{c}_2, \dots, \mathbf{c}_N \} \in \mathbb{R}^{N \times D}$, where $N$ is the codebook size, i.e., the total number of discrete tokens, and $D$ is the embedding dimensionality of each token vector. Each entry $\mathbf{c}_i \in \mathbb{R}^D$ corresponds to a learned prototype that represents a spatio-temporal video patch.

With temporal downsampling factor $d_t$ and spatial downsampling factor $d_s$, the tokenizer encodes the video into a grid of discrete tokens $\mathbf{Z} = \mathcal{E}(\mathbf{V}) \in \mathbb{R}^{t \times h \times w}$, where $t = \frac{T}{d_t}, \quad h = \frac{H}{d_s}, \quad w = \frac{W}{d_s}$. Here, $\mathbf{Z}$ is the vector of integer indices in the range $\{1,\dots,N\}$. By transmitting discrete tokens instead of raw pixels, the data volume is significantly reduced due to low-bit integer indexing, while preserving high-level semantics via learned codebook embeddings.

\subsection{Multi-modal User-intended Token Extraction}

User-intended semantic communication~\cite{Diff_Xinkai, gaze2023adaptive} has emerged as an influential research direction, driven by the need to align transmission priorities with user perception, attention, or task-specific goals rather than uniformly treating all content. In this work, we instantiate this principle in a multi-modal setting by extracting user-intended tokens from video content conditioned on textual intent, enabling differentiated processing and transmission at the token level.
\subsubsection{Text-conditioned Heatmap Generation} 
In order to find the related content in the input video according to the textual intent, for the first frame $\mathbf{I}_0 \in \mathbb{R}^{H \times W}$ in the sequence, we compute a vision-language model based text-conditioned heatmap for a textual intent description $\mathbf{p}$. The vision-language model patch size is $p$ (e.g., $p{=}32$ for ViT-B/32). The text-conditioned heatmap produced on the grid is
\begin{equation}
\hat{\mathbf{H}}\in\mathbb{R}^{\frac{H}{p}\times \frac{W}{p}},\qquad
\hat{\mathbf{H}}_{i,j}=\cos\!\big(f_{\text{img}}(\mathbf{I}_0^{(i,j)}),\,f_{\text{text}}(\mathbf{p})\big),
\end{equation}
where $\mathbf{I}_{0}^{(i,j)}$ denotes the $p\times p$ patch at grid index $(i,j)$ and $f_{\text{img}}, f_{\text{text}}$ are vision-language model's image and text encoders, and $\cos(\cdot, \cdot)$ denotes cosine similarity. We then normalize the heatmap $\hat{\mathbf{H}}_{i,j}$
\begin{equation}
\tilde{\mathbf{H}}_{i,j}=\frac{\hat{\mathbf{H}}_{i,j}-\min(\hat{\mathbf{H}})}{\max(\hat{\mathbf{H}})-\min(\hat{\mathbf{H}})+\varepsilon}\in[0,1],
\end{equation}
where $\varepsilon$ is a small constant to avoid division by zero.

To identify text-related regions, we apply a similarity threshold $\ell$ to the normalized heatmap. Specifically, a patch is selected as user-intended if its CLIP similarity exceeds $\ell$, yielding the binary patch-level mask
\begin{equation}
\mathbf{M}_0^{\text{patch}}(i,j)
= \mathbb{1}\!\big[ \tilde{\mathbf{H}}_{i,j} > \ell \big],
\end{equation}
where $\mathbb{1}[\cdot]$ is the indicator function. Here, $(h_p,w_p)=(H/p,\,W/p)$ denote the spatial resolution of the patch-level
heatmap. The resulting proportion of selected patches, denoted by $\rho$, is given by
\begin{equation}
\label{eq:rho_def}
\rho
=\frac{1}{h_pw_p}
\sum_{i=1}^{h_p}\sum_{j=1}^{w_p}
\mathbf{M}_0^{\text{patch}}(i,j),
\end{equation}
where the patch-level mask can optionally be refined by morphological dilation to enlarge high-response regions. Finally, the patch-level mask is upsampled to pixel resolution
\begin{equation}
\mathbf{M}_0
= \mathcal{U}\!\big(\mathbf{M}_0^{\text{patch}};\, H, W \big)
\in \{0,1\}^{H \times W},
\end{equation}
where $\mathcal{U}(\cdot)$ denotes nearest-neighbor upsampling, which replicates each patch label to its corresponding $p \times p$ pixel block.

\subsubsection{Dynamic Optical Flow Propagation}
We initialize the pixel-level binary semantic mask sequence $\{\mathbf{M}_{k}\}_{k=0}^{T-1}$ by setting the first frame $\mathbf{M}_0$. Let $\mathbf{F}_{k \rightarrow k+1} \in \mathbb{R}^{H \times W \times 2}$ denote the forward optical flow ~\cite{jampani2017video, weng2023mask} from frame $k$ to $k{+}1$, where each vector $\mathbf{F}_{k \rightarrow k+1}(u,v) = (\Delta x, \Delta y)$ is a pixel displacement, $(x,y)$ enumerates pixel coordinates and $(\Delta x,\Delta y)$ are taken from $\mathbf{F}_{k \rightarrow k+1}(y,x)$. We propagate the mask by warping with bilinear sampling $\mathcal{W}$
\begin{equation}\label{prop}
\mathbf{M}_{k+1} \;=\; \mathcal{W}\big(\mathbf{M}_{k}, \!\! \mathbf{F}_{k \rightarrow k+1}\big),
\quad
(x',y')=(x+\Delta x,\,y+\Delta y),
\end{equation}
where $k$ indexes original video frames ($k \in \{0,\dots,T{-}1\}$), $\mathcal{W}$ is a bilinear sampler, for each pixel, the flow 
displacement $(\Delta x,\Delta y)$ maps it to subpixel coordinates 
$(x',y')$, and the value is obtained by bilinear interpolation from 
the four neighboring pixels of $\mathbf{M}_k$. 

\subsubsection{Content Discrete Token Mapping}
\begin{figure}[t]
    \centering
    \includegraphics[width=0.5\textwidth]{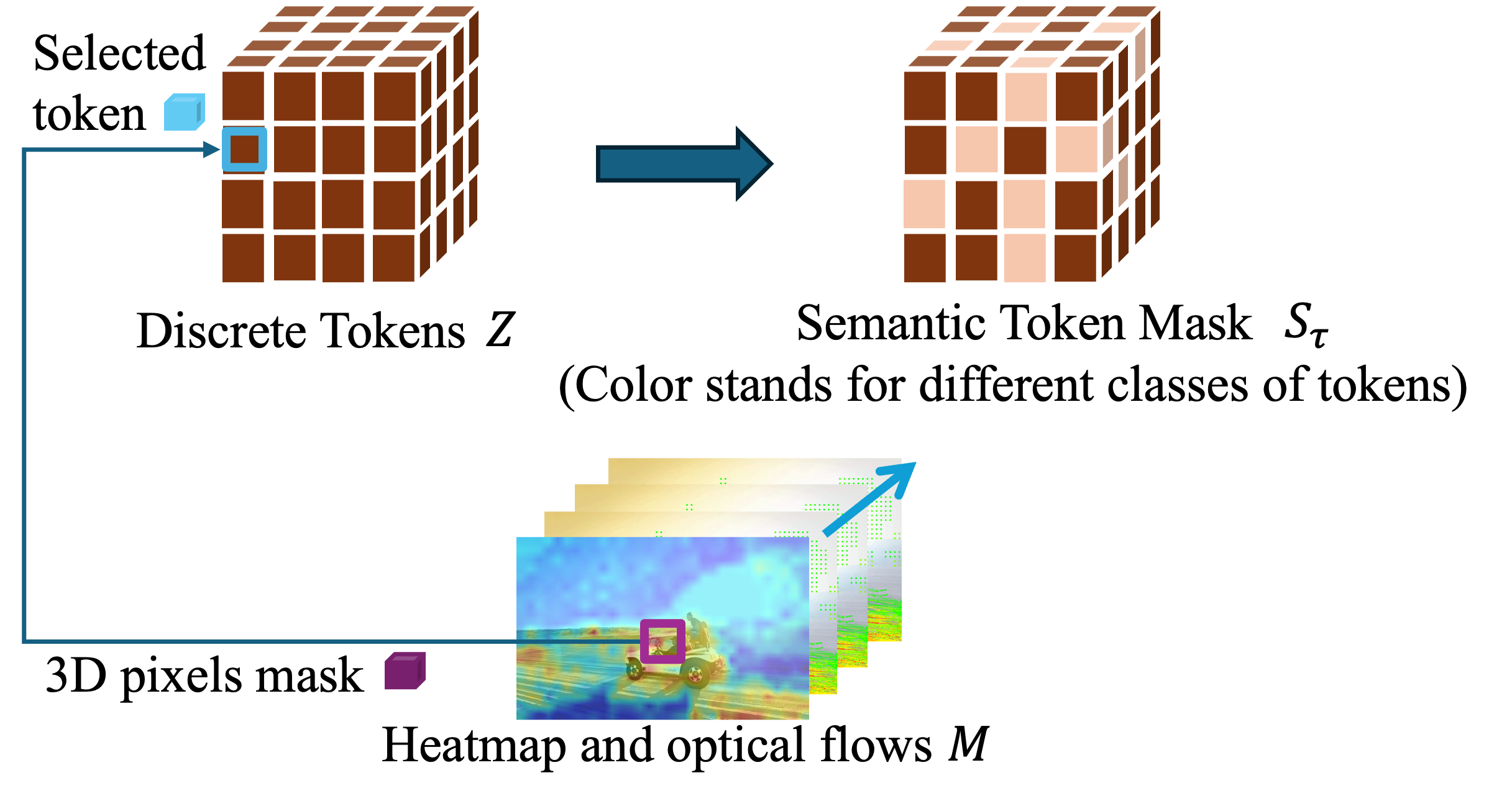} 
    \caption{Textual intent guided multiclass multirate token mapping.}
    \label{pic:mapping} 
\end{figure}

We partition all video tokens into two disjoint token classes $c\in\{s,n\}$, corresponding to user-intended tokens ($s$) and non-intended tokens ($n$), respectively. This classification is performed at the token level by mapping text-related content pixels to the discrete spatio-temporal token grid, as illustrated in Fig.~\ref{pic:mapping}. In order to map the text-related content pixels to discrete video tokens, we first note that the token temporal length is $t = T/d_t$ given the temporal downsampling factor $d_t$. Let $\tau \in \{0, \dots, t-1\}$ index the temporal tokens. The propagated pixel-level content mask at frame $\tau \cdot d_t + f$ is denoted by $\mathbf{M}_{\tau \cdot d_t + f} \in \{0,1\}^{H \times W}$, where $f \in \{0,\ldots,d_t-1\}$ indexes the original video frames within the temporal window corresponding to the $\tau$-th token. The spatio-temporal pooling aligns the propagated pixel-level masks with the
3D token grid
\begin{equation}\label{pooling}
\bar{\mathbf{M}}_{\tau} = \frac{1}{d_t} \sum_{f=0}^{d_t-1} \text{AdaptiveAvgPool2D}\big(\mathbf{M}_{\tau \cdot d_t + f}, (h, w)\big),
\end{equation}
where $\bar{\mathbf{M}}_{\tau}(i,j)\in[0,1]$ represents the average fraction of text-related pixels mapped to token $(i,j)$ over a temporal window of $d_t$ frames. A token-level binary semantic mask is obtained by thresholding
\begin{equation}
\mathbf{S}_{\tau} = \mathbb{1}\big[\bar{\mathbf{M}}_{\tau} > \theta\big],
\end{equation}
where $\mathbf{S}_{\tau} \in \{0,1\}^{h \times w}$ denotes the binary semantic token mask and $\theta \in (0,1]$ controls the minimum required proportion of text-related pixels for a token to be classified as user-intended. If the average proportion of text-related pixels within a spatio-temporal token, as captured by $\bar{\mathbf{M}}_{\tau}(i,j)$, exceeds the threshold $\theta$, that token is marked as user-intended.

A full description of the proposed multi-modal user-intended token extraction is provided in \textbf{Algorithm~\ref{alg:maskgen}}.

\begin{algorithm}[t]
\caption{Token-Level Semantic Extraction and Mask Generation}
\label{alg:maskgen}
\small
\KwInput{Video frames $\{\mathbf{I}_k\}_{k=0}^{T-1}$, textual intent $\mathbf{p}$.}
\KwOutput{Token-level semantic mask $\mathbf{S}=\{\mathbf{S}_\tau\}$; intended set $\mathcal{S}$ and non-intended set $\mathcal{N}$.}

Compute a vision--language heatmap on the first frame $\mathbf{I}_0$ and obtain
the initial pixel-level semantic mask $\mathbf{M}^{\rm(px)}_0$.

\For{$k=0$ \KwTo $T-2$}{
Propagate pixel-level semantic regions using optical flow:
\[
\mathbf{M}^{\rm(px)}_{k+1}
=\mathrm{Warp}\!\left(\mathbf{M}^{\rm(px)}_k,\mathbf{F}_{k\rightarrow k+1}\right).
\]
}

Downsample $\{\mathbf{M}^{\rm(px)}_k\}_{k=0}^{T-1}$ to the spatio-temporal token grid
and obtain the token-level semantic mask
$\mathbf{S}=\{\mathbf{S}_\tau\}_{\tau=0}^{t-1}$.

Define intended and non-intended token sets:
\[
\mathcal{S}=\{(\tau,i,j):\mathbf{S}_\tau(i,j)=1\}, \qquad
\mathcal{N}=\{(\tau,i,j):\mathbf{S}_\tau(i,j)=0\}.
\]

\textbf{Return} $\mathbf{S}$, $\mathcal{S}$, and $\mathcal{N}$.
\end{algorithm}

\subsection{Semantic-aware Multi-rate Bit Coding}

Given the discrete token index map $\mathbf{Z}=\{\mathbf{Z}_\tau\}$ produced by the tokenizer and the obtained semantic token masks $\mathbf{S}_{\tau}$, we allocate bit-precision according to semantic importance. Tokens inside the user-intended region are transmitted with the full codebook precision, while tokens outside the region are encoded using a reduced codebook defined in the token-index difference space. This achieves digital semantic communication ~\cite{huang2025d, latency_Li, oh2025blind, Chunmei_JSAC} with reduced bitrate while preserving semantically important content.

\paragraph{Full Codebook Precision Encoding of Intended Tokens}
For user-intended tokens where $\mathbf{S}_{\tau}[i,j]=1$, $B_{\text{full}}=\lceil\log_2 N\rceil$ is the full bit-precision for a token index (e.g., $B_{\text{full}}{=}16$ for 64000 codebook), therefore, the user-intended area tokens $\mathbf{Z}^{\text{intended}}_{\tau}[i,j]$ are transmitted in $B_{\text{full}}$ bits. 

\paragraph{Reduced Codebook Precision Differential Encoding of Non-Intended Tokens}
For non-user-intended tokens, we employ differential coding with a reduced precision $B_{\Delta}$. Let $\mathbf{Z}_{\text{ref}}$ denote the reference token map taken from the first frame $\mathbf{I}_0$. We define the raw differential value
\begin{equation}
x = \mathbf{Z}^{\text{non}}_{\tau}[i,j] - \mathbf{Z}_{\text{ref}}[i,j].
\end{equation}

A $B_{\Delta}$-bit signed quantizer can represent integer values within
\begin{equation}
-Q \le x \le Q,
\qquad
Q = 2^{B_{\Delta}-1}-1,
\end{equation}
which yields a total of $2Q+1 = 2^{B_{\Delta}}-1$ representable symbols. The valid representational set $\{-Q,\,\ldots,\,0,\,\ldots,\,Q\}$ constitutes a reduced codebook in the token-index difference space, containing $2Q+1 = 2^{B_\Delta}-1$ effective differential symbols within a $2^{B_\Delta}$-level unsigned representation. To ensure the differential value lies in this domain, we apply the symmetric clipping operator
\begin{equation}
\label{eq:clipping}
\Delta_{\tau}[i,j] = \mathcal{C}_Q(x)
= \min\!\big(\max(x,-Q), Q\big),
\end{equation}
where $\mathcal{C}_Q(\cdot)$ clips the difference to the symmetric range $[-Q, Q]$. 
Since bitstreams cannot directly store signed integers without additional sign handling, we shift this range to the non-negative interval by adding $Q$
\begin{equation}
\widetilde{\Delta}_{\tau}[i,j] \;=\; \Delta_{\tau}[i,j] + Q
\;\in\;\{0,\ldots,2Q\},
\end{equation}
the shifted differential values $\widetilde{\Delta}_{\tau}[i,j]$ are then transmitted using exactly $B_{\Delta}$ bits. This multi-rate bit-precision assignment enables the encoder to adapt the transmission rate to the semantic content. 

\paragraph{Token-level intended ratio and BPP analysis}
The patch-level intended ratio $\rho$ in~\eqref{eq:rho_def} represents the intent-related patches on the CLIP patch grid in the first frame. After optical-flow propagation~\eqref{prop}, and spatio-temporal pooling~\eqref{pooling}, intended/non-intended token classification is made on the video token spatio-temporal grid, $(h,w,t)=(H/d_s,\,W/d_s,\,T/d_t)$. The resulting token-level intended ratio is defined as
\begin{equation}
\label{eq:rhos_def}
\rho_s
=\frac{N_s}{t h w}
=\frac{1}{t h w}\sum_{\tau=0}^{t-1}\sum_{i=1}^{h}\sum_{j=1}^{w}\mathbf{S}_\tau(i,j),
\end{equation}
where $N_s$ is the number of user-intended tokens and $\mathbf{S}_\tau\in\{0,1\}^{h\times w}$ is the token-level semantic mask. Note that the resulting $\rho_s$ value depends on the CLIP threshold $\ell$ and the token-level threshold $\theta$. Given $\rho_s$, the total payload of bits for the video is
\begin{equation}
B_{\mathrm{tokens}}
= t h w\big(\rho_s B_{\mathrm{full}}+(1-\rho_s)B_{\Delta}\big),
\end{equation}
thereby normalizing by the video resolution $T\times H\times W$ over RGB channels, the resulting BPP value is
\begin{equation}
\mathrm{BPP}
=
\frac{1}{3 d_t d_s^2}
\big(\rho_s B_{\mathrm{full}}+(1-\rho_s)B_{\Delta}\big).
\end{equation}

A full description of the proposed semantic-aware multi-rate bit coding is provided in \textbf{Algorithm~\ref{alg:diffch}}.

\begin{algorithm}[t]
\caption{Semantic-aware Differential Token Encoding and Reconstruction}
\label{alg:diffch}
\small
\KwInput{Token indices $\mathbf{Z}=\{\mathbf{Z}_\tau\}_{\tau=0}^{t-1}$,
semantic token masks $\mathbf{S}=\{\mathbf{S}_\tau\}$,
bit-precisions $(B_{\text{full}},B_\Delta)$.}
\KwOutput{Encoded token stream $\mathbf{E}$ and reconstructed indices $\hat{\mathbf{Z}}$.}

\textbf{1. Reference token coding:}\\
Transmit the first token grid $\mathbf{Z}_{\text{ref}}$ in full precision:
\[
\mathbf{E}_{\text{ref}}=\mathbf{Z}_{\text{ref}}.
\]

\textbf{2. Differential token coding (for $\tau\ge1$):}
\For{$\tau=1$ \KwTo $t-1$}{
  \For{each token position $(i,j)$}{
    \eIf{$\mathbf{S}_\tau(i,j)=1$ \textnormal{(user-intended)}}{
      \[
      \mathbf{E}_\tau(i,j)=\mathbf{Z}_\tau(i,j)
      \quad (\text{use } B_{\text{full}} \text{ bits})
      \]
    }{
      Compute differential w.r.t. reference:
      \[
      x=\mathbf{Z}_\tau(i,j)-\mathbf{Z}_{\text{ref}}(i,j)
      \]
      Clip to $[-Q,Q]$, $Q=2^{B_\Delta-1}-1$:
      \[
      \Delta_\tau(i,j)=\mathcal{C}_Q(x)
      \]
      Shift and encode:
      \[
      \mathbf{E}_\tau(i,j)=\Delta_\tau(i,j)+Q
      \quad (\text{use } B_\Delta \text{ bits})
      \]
    }
  }
}
\textbf{3. Decoder reconstruction:}

Recover signed differentials for non-intended tokens:
\[
\widehat{\Delta}_{\tau}(i,j)
=
\hat{\mathbf{E}}_{\tau}(i,j) - Q,
\quad Q = 2^{B_\Delta-1}-1.
\]

Reconstruct token indices:
\[
\widehat{\mathbf{Z}}_{\tau}(i,j)=
\begin{cases}
\hat{\mathbf{E}}_{\tau}(i,j),
& \mathbf{S}_{\tau}(i,j)=1,\\[4pt]
\mathrm{clip}_{[0,N-1]}\!\big(
\mathbf{Z}_{\text{ref}}(i,j)+\widehat{\Delta}_{\tau}(i,j)
\big),
& \mathbf{S}_{\tau}(i,j)=0,
\end{cases}
\]
\textbf{Return} the encoded stream $\mathbf{E}$ and reconstructed tokens
$\widehat{\mathbf{Z}}=\{\widehat{\mathbf{Z}}_{\tau}\}$.

\end{algorithm}

\subsection{UEP-based Joint Distortion and Transmission Delay Minimization}

Let $\mathcal{T}$ denote temporal indices and $(i,j)$ spatial token locations. Semantic importance is derived from a vision-language related heatmap with optical-flow propagation, producing the intended-token set $\mathcal{S}\subseteq \mathcal{T}\times\mathbb{Z}^2$ and the non-intended set $\mathcal{N}$. All tokens in each class $c\in\{s,n\}$ share one transmission configuration selected from a finite candidate set $\mathcal{K}_c$.

\paragraph{UEP candidate sets}
For intended tokens, only the full semantic bit-precision $B_{\text{full}}$ is used, and the transmission configuration is chosen from
\[
\mathcal{K}_s=\{\text{QPSK}_{1/3},\; \text{QPSK}_{1/2}\}.
\]
For non-intended tokens, the encoder jointly chooses both the modulation--coding scheme and the differential bit-precision $B_\Delta$. Each candidate entry in $\mathcal{K}_n$ is of the form
\[
k = (B_\Delta^{(k)},\, m_k^{(n)},\, r_k^{(n)}),
\]
where $B_\Delta^{(k)}$ is selected from the reduced-precision set
\[
B_\Delta^{(k)} \in \mathcal{B}_n
= \{16,15,14,13,12,11,10\}.
\]
The pair $\big(m_k^{(n)}, r_k^{(n)}\big)$ is determined by the selected
modulation--coding scheme (MCS), with the candidate MCS set given by
\[
\mathcal{M}_n =
\{\text{QPSK}_{1/3},\; \text{QPSK}_{1/2},\;
\text{16QAM}_{1/2},\; \text{16QAM}_{3/4}\}.
\]
Here, $m_k^{(n)}$ denotes the modulation order (bits per symbol), and
$r_k^{(n)}$ is the channel coding rate. Each configuration is feasible only when the instantaneous SNR exceeds
its activation threshold.
\paragraph{PDU aggregation and per-candidate metrics}
Tokens are transmitted in protocol data units (PDUs). Let $L_c$ be the number of tokens per PDU and $H_c$ the PDU overhead in bits (e.g., headers, pilots, CRC). For $k\in\mathcal{K}_c$, with token bit-precision $B_k^{(c)}$, code rate $r_k^{(c)}$, and modulation order $m_k^{(c)}$, the spectral efficiency is
\begin{equation}
\label{eq:spec_eff}
g_{k,\mathrm{raw}}^{(c)} = \rho_{\mathrm{oh}}\, m_k^{(c)} r_k^{(c)},
\end{equation}
where $\rho_{\mathrm{oh}}\in(0,1)$ represents the effective physical layer resource utilization factor accounting for reference signals, control overhead, and other non-data resource elements. The BLER is tabulated as a function of the modulation order $m$ and code rate $r$, indexed by MCS values, similar to the 3GPP NR standard~\cite{3GPP_38_214, 3GPP_38_212}. Specifically, the corresponding BLER
is given by
\begin{equation}
p_k^{(c)} 
= \mathrm{BLER}\!\left(m_k^{(c)},\, r_k^{(c)} \mid \gamma \right),
\end{equation}
where $\gamma$ denotes the instantaneous SNR. In simulation, $p_k^{(c)}$ controls a PDU-level packet loss process in which tokens are grouped into PDUs of $L_c$ tokens and each PDU is lost with independent probability $p_k^{(c)}$. Tokens belonging to lost PDUs are replaced by the reference token $Z_{\mathrm{ref}}$ for reconstruction at the receiver.

A PDU carries $B_{\mathrm{PDU},k}^{(c)} = L_c B_k^{(c)} + H_c$ bits in total, including payload and overhead. The resource consumption (in Hz$\cdot$s) and transmission time (in s) per PDU are
\begin{equation}
\bar{C}_{\mathrm{PDU},k}^{(c)}
  = \frac{B_{\mathrm{PDU},k}^{(c)}}{g_{k,\mathrm{raw}}^{(c)}}, \qquad
\bar{T}_{\mathrm{PDU},k}^{(c)}
  = \frac{\bar{C}_{\mathrm{PDU},k}^{(c)}}{W_B},
\end{equation}
where $W_B$ is the system bandwidth in Hz.
The induced per-token resource and delay are obtained by averaging over
the $L_c$ tokens carried in a PDU
\begin{equation}
c_k^{(c)}=\frac{\bar{C}_{\mathrm{PDU},k}^{(c)}}{L_c},\qquad
\tau_k^{(c)}=\frac{\bar{T}_{\mathrm{PDU},k}^{(c)}}{L_c},
\end{equation}
where $c_k^{(c)}$ (in Hz$\cdot$s per token) and $\tau_k^{(c)}$ (in seconds per token) are exactly the scalar costs that enter the joint distortion--delay optimization.
Quantization distortion is approximated by an exponential model
\begin{equation}
\label{eq:rate_dist}
d_k^{(c)} = \alpha_c e^{-\beta_c B_k^{(c)}},
\end{equation}
which is motivated by the rate-distortion-perception theorem~\cite{ICML_Blau, JASC_Chen}, to model the convex and monotonically non-increasing behavior of distortion and perception metrics as a function of the rate, with $\alpha, \beta$ parameters to be determined empirically by curve fitting in experiments ~\cite{Diff_Xinkai}. 

\paragraph{Class-level decision variables}
Since all tokens within the same semantic class share one transmission configuration, each class $c\in\{s,n\}$ selects exactly one entry from its candidate set $\mathcal{K}_c$. This is encoded by the binary decision variables
\begin{equation}
z_k^{(c)}\in\{0,1\}, \qquad 
\sum_{k\in\mathcal{K}_c} z_k^{(c)} = 1,
\end{equation}
where $z_k^{(c)}=1$ indicates that configuration $k$ is chosen for every token belonging to class $c$.
Let $N_s=|\mathcal{S}|$ and $N_n=|\mathcal{N}|$ denote the number of tokens in the intended and non-intended sets, respectively. Because all tokens in class $c$ use the same configuration, the total distortion, delay, and resource consumption become linear functions of the selection variables. Specifically, the total distortion is
\begin{equation}
D_\mathrm{tot}
= N_s\!\sum_{k\in\mathcal{K}_s} d_k^{(s)} z_k^{(s)}
+ N_n\!\sum_{k\in\mathcal{K}_n} d_k^{(n)} z_k^{(n)},
\end{equation}
where $N_s$ copies of distortion $d_k^{(s)}$ are incurred when class $s$
chooses configuration $k$, and similarly for class $n$.
The total delay is
\begin{equation}
\tau_{\mathrm{tot}}
= N_s\!\sum_{k\in\mathcal{K}_s} \tau_k^{(s)} z_k^{(s)}
+ N_n\!\sum_{k\in\mathcal{K}_n} \tau_k^{(n)} z_k^{(n)}.
\end{equation}

\paragraph{Resource constraint} 
A fixed fraction $\eta\in(0,1]$ of the time–frequency resource block is made available for token transmission.
The resulting resource budget (in Hz$\cdot$s) is
\begin{equation}
R_{\max} = \eta\, W_B\, T_{\mathrm{RB}},
\end{equation}
in which $W_B$ and $T_{\mathrm{RB}}$ denote the frequency width and time length of each resource block. The total resource consumption in (Hz$\cdot$s) is
\begin{equation}
\label{eq:Rsched_def}
N_s\!\sum_{k\in\mathcal{K}_s} c_k^{(s)} z_k^{(s)}
+ N_n\!\sum_{k\in\mathcal{K}_n} c_k^{(n)} z_k^{(n)} \le R_{\max},
\end{equation}
which should be smaller than $ R_{\max}$. 

\paragraph{Weighted distortion–delay optimization}
Distortion and delay are normalized by their feasible extrema
\begin{equation}
D_{\mathrm{norm}}=\frac{D_\mathrm{tot}-D_{\min}}{D_{\max}-D_{\min}+\epsilon_D},
\qquad
T_{\mathrm{norm}}=\frac{\tau_{\mathrm{tot}}-T_{\min}}{T_{\max}-T_{\min}+\epsilon_T}.
\end{equation}
In the simulation, the normalization constants $\{D_{\min},D_{\max},T_{\min},T_{\max}\}$ are computed from class-wise extrema over the finite candidate sets.
Specifically, let $\underline{d}^{(c)}=\min_{k\in\mathcal{K}_c} d_k^{(c)}$ and $\overline{d}^{(c)}=\max_{k\in\mathcal{K}_c} d_k^{(c)}$, and similarly define $\underline{\tau}^{(c)}$ and $\overline{\tau}^{(c)}$ from $\{\tau_k^{(c)}\}_{k\in\mathcal{K}_c}$. We then set
\begin{equation}
D_{\min}=N_s\,\underline{d}^{(s)}+N_n\,\underline{d}^{(n)},\quad
D_{\max}=N_s\,\overline{d}^{(s)}+N_n\,\overline{d}^{(n)},
\end{equation}
\begin{equation}
T_{\min}=N_s\,\underline{\tau}^{(s)}+N_n\,\underline{\tau}^{(n)},\quad
T_{\max}=N_s\,\overline{\tau}^{(s)}+N_n\,\overline{\tau}^{(n)}.
\end{equation}
This class-wise construction yields numerically stable normalization while preserving the relative scale between distortion and delay. The optimization objective is

\begin{subequations}\label{eq:optMILP}
\begin{align}
\min_{\{z_k^{(c)}\}} \quad 
    & J = w_{\mathrm{D}} D_{\mathrm{norm}} + w_{\mathrm{T}} T_{\mathrm{norm}} \nonumber \\
\text{s.t.}\quad
    & N_s\!\sum_{k\in\mathcal{K}_s} c_k^{(s)} z_k^{(s)}
+ N_n\!\sum_{k\in\mathcal{K}_n} c_k^{(n)} z_k^{(n)} \le R_{\max}, 
      \label{eq:optMILP_a} \\
    & \sum_k z_k^{(c)} = 1,
      \label{eq:optMILP_b} \\
    & z_k^{(c)} \in \{0,1\},
      \label{eq:optMILP_c} \\
    & \sum_k p_k^{(s)} z_k^{(s)} \le p_{\max}^{(s)},
      \label{eq:optMILP_d} \\
    & \sum_k p_k^{(n)} z_k^{(n)} \le p_{\max}^{(n)}(\mathrm{SNR}),
      \label{eq:optMILP_e}
\end{align}
\end{subequations}
where $w_{\mathrm{D}}\ge 0$ and $w_{\mathrm{T}}\ge 0$ are weighting factors that
control the tradeoff between semantic distortion and transmission delay, with $w_{\mathrm{D}}+w_{\mathrm{T}}=1$. The BLER constraints ensure that the chosen operating point stays within a reliability region compatible with the instantaneous channel conditions, thereby controlling the packet error probability.
 
We propose \textbf{Algorithm \ref{alg:opt}} based on mixed integer linear programming (MILP)~\cite{NemhauserWolsey1988} to solve the optimization problem \eqref{eq:optMILP} for joint bit-precision and MCS selection. In general, solving a MILP is NP-hard and its worst-case complexity grows exponentially with the number of integer decision variables. In our formulation, we have the binary variables $\{z_k^{(s)}\}_{k\in\mathcal{K}_s}$ and $\{z_k^{(n)}\}_{k\in\mathcal{K}_n}$, 
and thus the number of binary variables is $N_{\mathrm{bin}} = |\mathcal{K}_s| + |\mathcal{K}_n|$. Since the candidate sets $\mathcal{K}_s$ and $\mathcal{K}_n$ are typically small and the MILP is solved once per scheduling window, the optimization remains computationally feasible in practice. 

\begin{algorithm}[t]
\caption{Joint Bit-Precision and MCS Adaptation}
\label{alg:opt}
\small
\KwInput{SNR $\gamma$, bandwidth $W_B$, overhead factor $\rho_{\mathrm{oh}}
$,
resource budget $R_{\max}$,
token counts $(N_s,N_n)$,
bit-precision sets $\mathcal{B}_s,\mathcal{B}_n$,
MCS sets $\mathcal{M}_s,\mathcal{M}_n$,
PDU packing parameters $\{(L_c,H_c)\}_{c\in\{s,n\}}$,
weights $(w_D,w_T)$.}
\KwOutput{$(b_S^\star,m_S^\star,b_N^\star,m_N^\star)$.}

\textbf{Candidate generation:}

\For{$c \in \{s,n\}$}{
    \If{$c=s$}{
        $\mathcal{B}_s \leftarrow \{B_{\text{full}}\}$.
    }
    \For{$(b,m)\in\mathcal{B}_c\times\mathcal{M}_c$}{
        \textit{(SNR feasibility)} keep $(b,m)$ only if $\gamma \ge \gamma_{\min}(m)$.

        Compute BLER from a link-level table:
        \[
        p_k^{(c)}=\mathrm{BLER}(m_k^{(c)}, r_k^{(c)}\mid\gamma).
        \]

        Spectral efficiency:
        \[
        g_{k,\mathrm{raw}}^{(c)}=\rho_{\mathrm{oh}}\, m_{\rm mod}(m)\, r_m.
        \]

        PDU bits (payload + overhead): $B_{\rm PDU}=L_c\,b+H_c$.

        Per-token resource cost:
        $c_k^{(c)}=\frac{B_{\rm PDU}}{L_c\, g_{k,\mathrm{raw}}^{(c)}}$.

        Per-token delay: $\tau_k^{(c)}=\frac{c_k^{(c)}}{W_B}$.

        Distortion proxy: $d=\alpha_c \exp(-\beta_c b)$.

        Store $(c_k^{(c)},\tau_k^{(c)},d,p_k^{(c)},b,m)$ in the class-$c$ candidate list.
    }
}

\textbf{MILP solution:} Solve the joint bit-precision and MCS selection problem
to obtain the optimal configurations $(b_S^\star,m_S^\star)$ and
$(b_N^\star,m_N^\star)$.

\textbf{Return} $(b_S^\star,m_S^\star,b_N^\star,m_N^\star)$.
\end{algorithm}

\subsection{Token-based Reconstruction}
At the receiver, for user-intended tokens, read the transmitted full value, and we recover non-user-intended area tokens by inverse offset and addition to
the reference frame's tokens
\begin{equation}
\widehat{\Delta}_{\tau}[i,j] \;=\; \widetilde{\Delta}_{\tau}[i,j] - Q, 
\end{equation}
\begin{equation}
\widehat{\mathbf{Z}}_{\tau}[i,j] \;=\;
\begin{cases}
\mathbf{Z}^{\text{intended}}_{\tau}[i,j], & \mathbf{S}_{\tau}[i,j]=1,\\[2pt]
\mathrm{clip}_{[0,N{-}1]}\!\big(\mathbf{Z}_{\text{ref}}[i,j]+\widehat{\Delta}_{\tau}[i,j]\big),
& \mathbf{S}_{\tau}[i,j]=0.
\end{cases}
\end{equation}
Here, $\mathrm{clip}_{[0,N-1]}(\cdot)$, as shown in equation~\eqref{eq:clipping}, denotes the clipping operator that restricts reconstructed token indices to the valid codebook range. User-intended tokens are directly transmitted in full precision, while non-user-intended tokens are reconstructed from their $B_\Delta$-bit differential representation, obtained by clipping the raw difference to $[-Q,Q]$ and transmitting its $Q$-shifted quantized form. Finally, we decode the discrete tokens using the tokenizer's decoder $\mathcal{D}(\cdot)$
\begin{equation}
\hat{\mathbf{V}} = \mathcal{D}(\widehat{\mathbf{Z}}_{\tau}),
\end{equation}
where $\hat{\mathbf{V}} \in \mathbb{R}^{T \times H \times W}$ is the reconstructed video.

At the decoder, the token-level semantic mask $\mathbf{S}_\tau$ is required to distinguish tokens reconstructed from full indices and from differentials.
Rather than transmitting $\mathbf{S}_\tau$ for all frames, we signal only the
first-frame mask and propagate it temporally using lightweight block-level
motion vectors on the token grid. For a token grid of size $t\times h\times w$ and a block size of $g\times g$, the resulting relative side-information
overhead is
\begin{equation}
\label{eq:overhead}
\frac{B_{\mathrm{side}}}{B_{\mathrm{tokens}}}
=
\frac{\tfrac{1}{t}+\tfrac{(t-1)}{t}\tfrac{B_{\mathrm{mv}}}{g^2}}
{\rho_s B_{\mathrm{full}} + (1-\rho_s) B_\Delta},
\end{equation}
where $\rho_s = N_s/(t h w)$ is the fraction of user-intended tokens. With small coarse blocks $g$ and compact motion quantization $B_{\mathrm{mv}}$, the side information typically accounts for only a few percent of the token payload.

\begin{table*}[t]
\centering
\caption{Performance comparison between the proposed video TokenCom framework (BPP=0.013), with VC-DM~\cite{videocompress_bench} and H.265~\cite{ISO23008-2:2025} (BPP=0.02) baselines on the UVG dataset.}
\label{tab:uvg_comparison}
\renewcommand{\arraystretch}{1.25}
\begin{tabular}{l|ccc|ccc|ccc}
\hline
\textbf{UVG Video Sample} &
\multicolumn{3}{c|}{\textbf{PSNR}} &
\multicolumn{3}{c|}{\textbf{LPIPS}} &
\multicolumn{3}{c}{\textbf{FVD}} \\
\cline{2-10}
 & TokenCom & VC-DM~\cite{videocompress_bench} & H.265~\cite{ISO23008-2:2025}
 & TokenCom & VC-DM~\cite{videocompress_bench} & H.265~\cite{ISO23008-2:2025}
 & TokenCom & VC-DM~\cite{videocompress_bench} & H.265~\cite{ISO23008-2:2025} \\
\hline
UVG-ShakeNDry      & \textbf{26.06} & 24.68 & 24.43 & 0.1296 & \textbf{0.111} & 0.158 & \textbf{1076} & 1400 & 2896 \\
UVG-Beauty         & \textbf{30.57} & 28.45 & 25.28 & \textbf{0.0469} & 0.057 & 0.17 & \textbf{554}  & 1416 &3227 \\
UVG-Bosphorus      & \textbf{28.94} & 24.55 & 26.70 & \textbf{0.0863} & 0.101 & 0.104 & \textbf{1406} & 2079 & 2951 \\
UVG-HoneyBee       & \textbf{22.49} & ---  & 21.55 & \textbf{0.1683} & ---   & 0.196 & \textbf{504}  & --- & 1446 \\
UVG-Jockey         & \textbf{24.31} & 22.82 & 20.95 & \textbf{0.0933} & 0.147 & 0.201 & \textbf{965}  & 2349 & 6426 \\
UVG-ReadySteadyGo  & \textbf{22.38} & 20.70 & 20.11 & 0.1307 & \textbf{0.112} & 0.316 & \textbf{2213} & 2832 & 6347 \\
UVG-YachtRide & \textbf{25.94} & 25.61 & 23.95 & \textbf{0.087}  & 0.096 & 0.14 & \textbf{1522} & 2540 & 4282 \\
\hline
\textbf{Average}
 & \textbf{26.36} & 24.47 & 23.28
 & \textbf{0.095} & 0.104 & 0.184
 & \textbf{1289}  & 2087 & 4010 \\
\hline
\end{tabular}
\end{table*}

\begin{figure*}[t]
    \centering
    \includegraphics[width=0.75\linewidth]{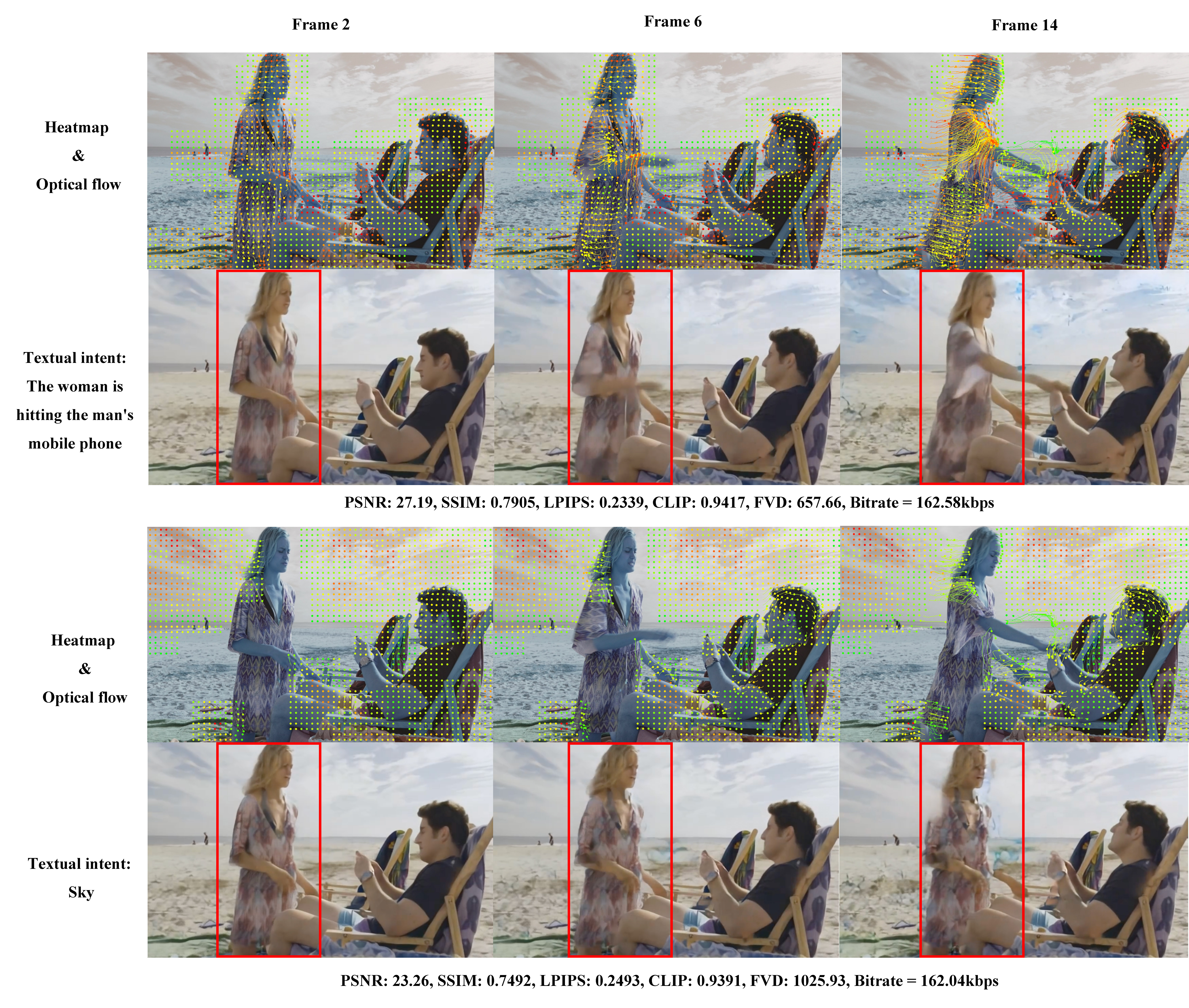} 
    \caption{Performance of the proposed video TokenCom framework with different textual intents: ``The woman is hitting the man's mobile phone.'' and ``Sky.''. The bit-precision for transmitting user intended and non-intended tokens is 16 and 11 bits per token, respectively. The red rectangle shows the user-intended regions guided by the textual intent.}
    \label{fig:women_hit_1} 
\end{figure*}

\begin{figure*}[t]
    \centering
    \includegraphics[width=0.8\linewidth]{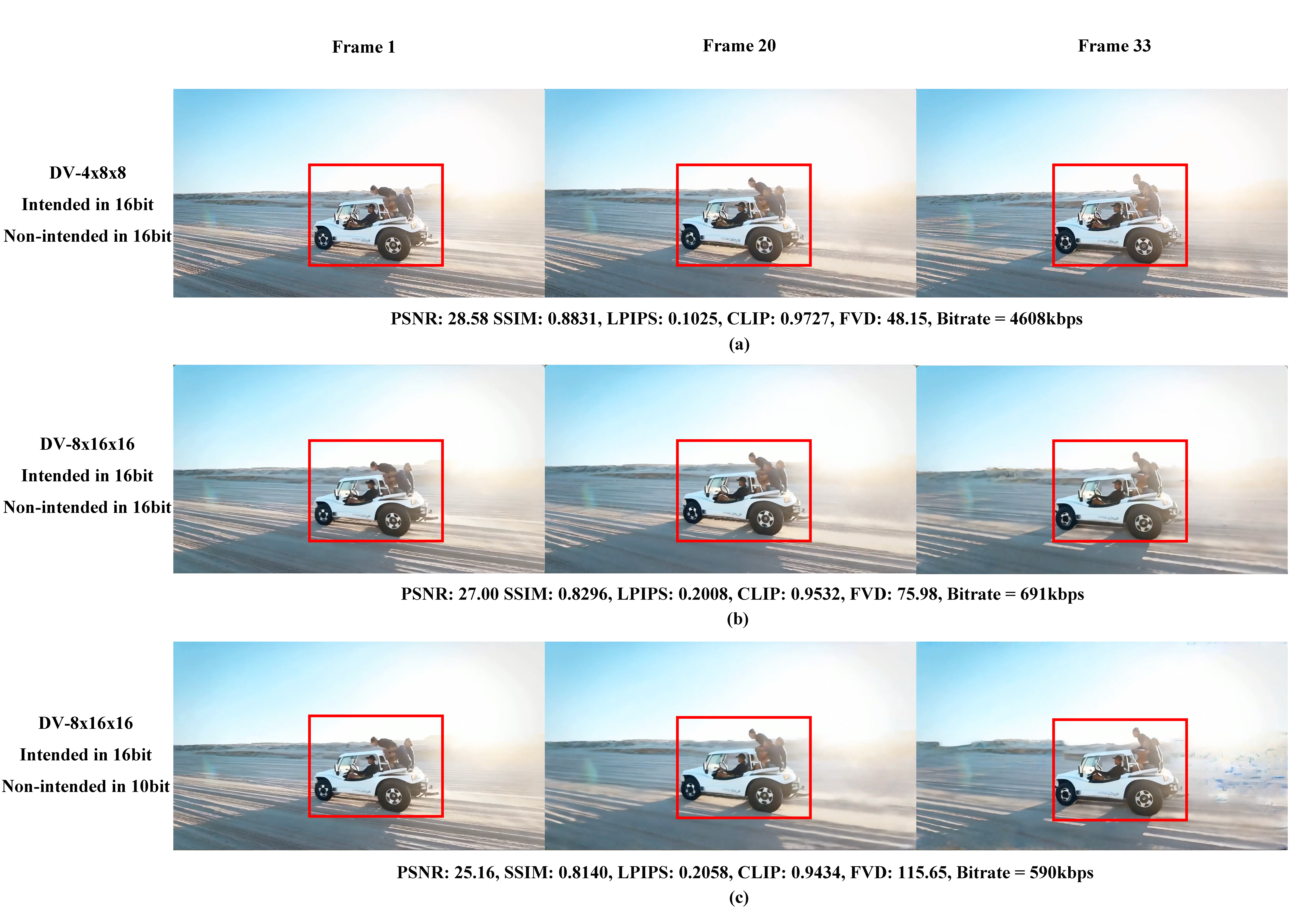} 
    \caption{Performance of the proposed video TokenCom framework. The red rectangle shows the user-intended regions guided by the textual intent ``Car and person.''}
    \label{fig:car_desert_2} 
\end{figure*}

\section{Experimental Results}
For the dataset, we selected the widely used MCL-JCV~\cite{MCL-JCV} and UVG~\cite{UVG} video datasets in YUV format. Our MCL-JCV subset is cropped to a resolution of 1024$\times$640. For the UVG dataset, we use the videos whose native resolution is 1920$\times$1080. The main objects appearing in each video are captioned as the user-intended text description, serving as the textual intent input to the system. In this simulation, we employ the pre-trained Cosmos~\cite{nvidia2024cosmos} DV-$8\times16\times16$ and DV-$4\times8\times8$ models as the video tokenizer, and pre-trained CLIP~\cite{radford2021clip} as the vision–language model for semantic relevance estimation. It is worth noting that CLIP is adopted here as a representative vision–language model, and can be readily replaced by other pretrained vision–language models without affecting the proposed framework, since the semantic token extraction only relies on a generic image–text similarity score. Moreover, the proposed framework is agnostic to the specific choice of tokenizer and can accommodate any discrete video tokenization model that produces a spatio-temporal grid of token indices. To balance the rate and distortion, we fix the CLIP similarity threshold to $\ell=0.6$ in Section~II-B1 and the token-level threshold to $\theta=0.3$ in Section~II-B3. In addition, the physical layer resource utilization factor is fixed to $\rho_{\mathrm{oh}}=0.85$ in equation~\eqref{eq:spec_eff}. The parameters $\alpha$ and $\beta$ in equation~\eqref{eq:rate_dist} are determined empirically via curve fitting, and we use $\alpha_c = 1.0$ and $\beta_c = 0.2$ throughout our experiments. For a balanced trade-off between the resulting distortion and delay, we set the weighting factors to $w_{\mathrm{D}} = 0.5$ and $w_{\mathrm{T}} = 0.5$ in equation~\eqref{eq:optMILP}.

\subsection{Comparison with baselines}
We compare our method with a state-of-the-art diffusion-based ultra-low-bitrate generative SemCom scheme VC-DM~\cite{videocompress_bench}, as well as the conventional widely adopted H.265~\cite{hevc, ISO23008-2:2025} digital video compression codec. As shown in Table~\ref{tab:uvg_comparison}, VC-DM is trained and evaluated on UVG videos cropped to a spatial resolution of 128$\times$128 with a temporal length of 30 frames. For a fair comparison, we therefore downsample and crop the UVG dataset to the same resolution and number of frames, and adopt the identical PSNR, LPIPS and FVD computation protocol used in VC-DM. The results in Table~\ref{tab:uvg_comparison} indicate that our proposed framework consistently outperforms VC-DM and H.265 across all test videos. Moreover, our method achieves a lower rate of 0.013\,BPP compared with 0.02\,BPP for VC-DM and H.265, further demonstrating the superior rate--distortion efficiency of the proposed Video TokenCom framework.

\subsection{Textual Intent-based Source Coding}
First, we evaluated the controllability of different user-intended texts. We visualized the results using heatmaps and optical flow maps. In the heatmap and optical flow in Fig.~\ref{fig:women_hit_1}, red and orange dots indicate pixel regions highly correlated with the user's intent, yellow dots indicate medium correlation, while green corresponds to low correlation. From the results in Fig.~\ref{fig:women_hit_1}, when the user-intended text was \textit{``The woman is hitting the man's mobile phone''}, the optical flow in the heatmap was more concentrated on the \textit{woman}, \textit{man}, \textit{hitting}, and \textit{mobile phone} regions, related to the user-intended text. The bit-precision for transmitting user intended tokens is 16 bit per token, while the non-intended tokens are transmitted with 11 bits per token from a differentially reduced codebook. Moreover, the reconstruction quality for the woman in the red bounding box is improved compared to using \textit{``Sky''} as the intended text, and the overall performance metrics of the video is higher. Conversely, when the user-intended text is \textit{``Sky''}, the semantic importance shifts towards sky-related regions, leading to higher reconstruction quality in the sky area, while less relevant regions receive reduced bit allocation. Despite different intents, the overall bitrate remains similar in both cases, approximately 160~kbps, since the two textual intents cover a comparable number of pixels. This shows that our proposed method can keep the reconstruction quality of user-intended area according to different textual intent.

In Fig.~\ref{fig:car_desert_2}, to demonstrate performance of our proposed textual intent-guided video TokenCom framework on longer videos, we evaluate different bit-precisions on a 400 frames video at 1920$\times$1080 resolution from Pexels~\cite{pexels}. In Fig.~\ref{fig:car_desert_2} (a) we use the high quality Cosmos tokenizer DV-4$\times$8$\times$8, which gives a bitrate of 4608 Kbps. If the channel/network condition deteriorates significantly, the proposed TokenCom framework can reduce the transmission bitrate to 691 Kbps by switching to a lower quality tokenizer, i.e. Cosmos tokenizer DV-8$\times$16$\times$16 here. If the channel/network condition further deteriorates, textual intent awareness enables our proposed framework to reduce the bitrate even further down to 590 Kbps, without losing quality in the intended regions, as guided by the textual intent, i.e. \textit{``Car and person''}. This result demonstrates the significant adaptation capability of our proposed framework, either by changing the tokenizer, or by using the textual intent when the tokenizer remains fixed.

\subsection{Channel-Adaptive Video TokenCom}
\begin{figure}[t]
    \centering
    \includegraphics[width=0.5\textwidth]{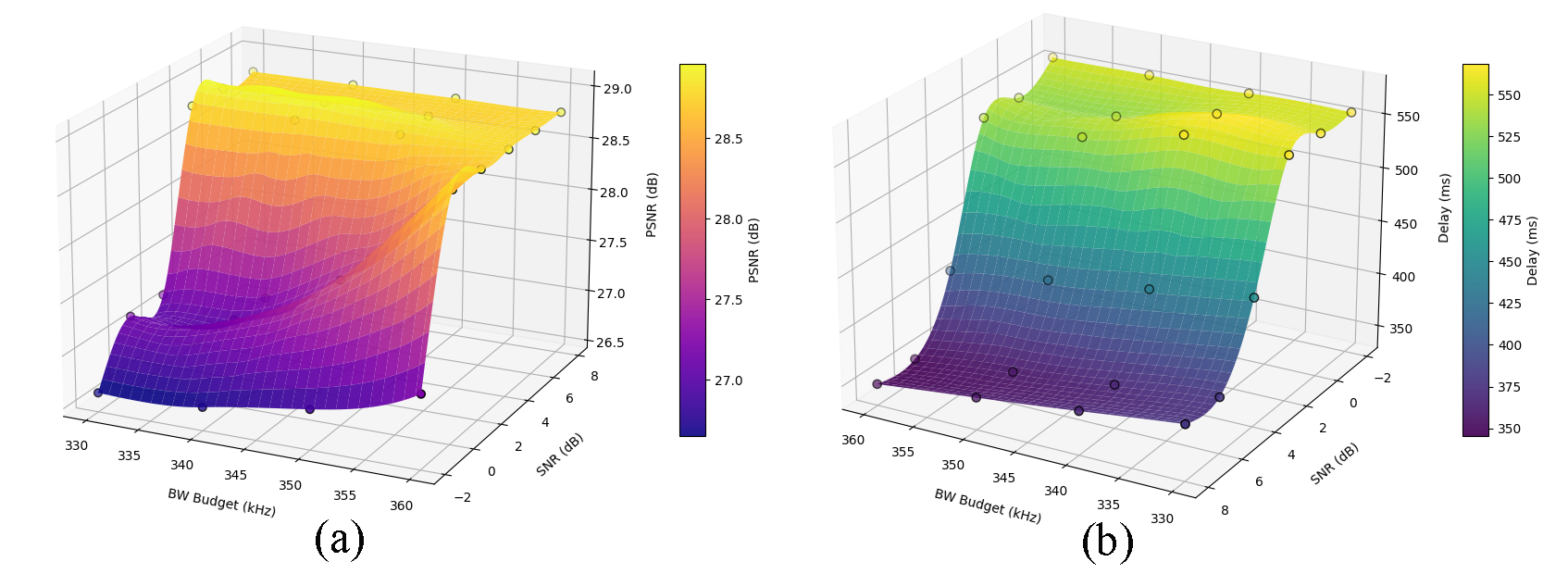} 
    \caption{(a) The achievable PSNR surface for source and channel coding distortion optimization; (b) The achievable delay surface for source and channel coding adaptation, by the proposed video TokenCom framework.}
    \label{pic:optimize} 
\end{figure}

To demonstrate the effectiveness of our joint adaptive source–channel coding scheme, we further analyze how the distortion and latency evolve under different bandwidth (BW) budgets, to achieve latency-aware SemCom ~\cite{latency_Li, Low_latency_Chong}. As Fig.~\ref{pic:optimize}(a) illustrates, we evaluate the \textit{``The woman is hitting the man's mobile phone''} video from the MCL-JCV dataset under four bandwidth settings, namely $330\mathrm{kHz}$, $340\mathrm{kHz}$, $350\mathrm{kHz}$, and $360\mathrm{kHz}$, while keeping the time--frequency resource block size fixed. For a fixed SNR, the PSNR monotonically improves as the bandwidth budget increases. This occurs because a larger bandwidth allows the optimizer to select higher bit-precision and stronger MCS levels, thereby reducing quantization distortion and improving reconstruction quality. In addition, Fig.~\ref{pic:optimize}(b) reports the corresponding delay. For a fixed SNR, increasing the bandwidth budget generally reduces the end-to-end transmission delay. This is because the throughput gain provided by a larger bandwidth outweighs the moderate increase in payload size introduced by higher source bit-precision allocation.

\begin{figure*}[t]
    \centering
    \includegraphics[width=0.9\linewidth]{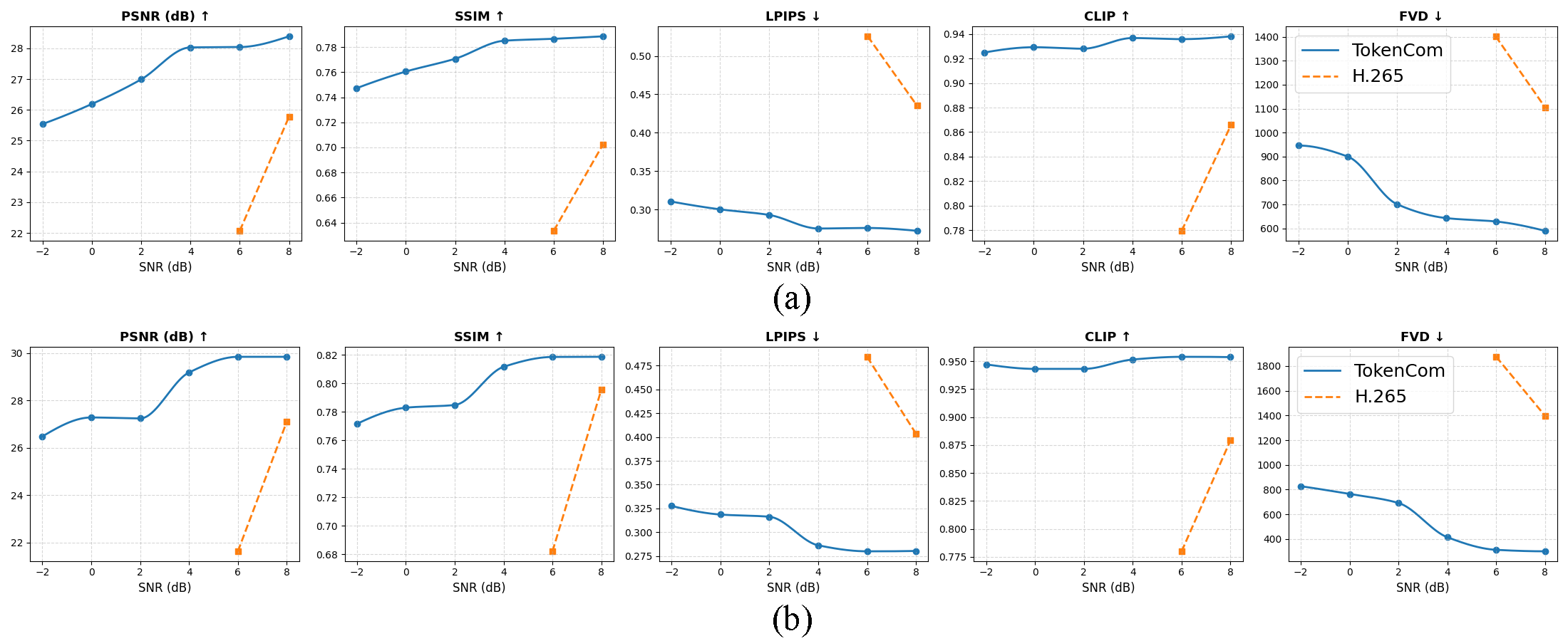} 
    \caption{The comparison results with benchmark under the MCL-JCV (a) and UVG (b) datasets. Note that some H.265 points are not shown at low SNRs because the adaptive H.265 pipeline frequently ``Failed" to decode more than 85\% of the frames, and such cases are marked as invalid in our evaluation.}
    \label{fig:curve_h265} 
\end{figure*}

\begin{table}[t]
\centering
\caption{Comparison of Key Parameters Between the Proposed video TokenCom Framework and H.265}
\label{tab:parameter_compare}
\begin{tabular}{p{2.4cm} p{2.4cm} p{2.4cm}}
\hline

\textbf{ } & \textbf{TokenCom (Proposed)} & \textbf{H.265 (Baseline)} \\
\hline
Information unit 
    & Discrete video tokens (codebook indices)
    & Pixel-domain blocks (CTU/CU/TU) \\

Coding granularity 
    & Token-level full-bit / differential-bit
    & Block-level (16--64 px) \\

Source coding 
    & Multi-rate token coding (full-precision)
    & Transform + prediction + CABAC \\

Semantic awareness 
    & Explicit (CLIP heatmap + optical-flow propagation)
    & None \\

Rate adaptation 
    & Per-token intended vs non-intended bit allocation
    & QP-based constant bitrate mode; uniform per frame/slice \\

Unequal error protection
    & Semantic UEP (separate MCS for token classes)
    & No UEP \\

Transmission unit 
    & Token bits also packed into PDUs
    & Bitstream segmented into PDUs \\

PDU size
    & Adapts from 1024 to 1280 bytes
    & 1200 bytes \\

Resource budget 
    & Total $B\!\times\!T_{\text{RB}}$ 
    & Total $B\!\times\!T_{\text{RB}}$ \\

Bandwidth in experiments 
    & 350\,kHz (MCL-JCV), 1060\,kHz (UVG) 
    & 350\,kHz (MCL-JCV), 1060\,kHz (UVG) \\
\hline
\end{tabular}
\end{table}

\begin{table*}[t]
\centering
\caption{Side-by-Side Comparison of the proposed video TokenCom framework and H.265 Baseline for a typical video. \\(``Failed" means that less than 85\% of the frames could be reconstructed successfully.)}
\label{tab:optimized_compare}
\scriptsize

\begin{tabular}{c@{\hspace{0.5cm}}c}


\begin{minipage}{0.46\linewidth}
\centering
\textbf{(a) TokenCom (Proposed)} \\[4pt]

\begin{tabular}{|c|c|c|c|c|c|c|c|c|}
\hline
\textbf{SNR} 
& \textbf{Mod$_{int}$} & \textbf{R$_{int}$}
& \textbf{Mod$_{non}$} & \textbf{R$_{non}$}
& \boldmath$B_{\Delta}$
& \textbf{Source Bitrate}
& \textbf{PSNR} \\
\hline

-2 & QPSK & 0.333 & QPSK & 0.500 & 12 & 165 & 26.69 \\ \hline
0  & QPSK & 0.333 & QPSK & 0.500 & 12 & 165 & 27.40 \\ \hline
2  & QPSK & 0.333 & QPSK & 0.500 & 12 & 165 & 27.40 \\ \hline
4  & QPSK & 0.500 & QPSK & 0.500 & 16 & 173 & 28.82 \\ \hline
6  & QPSK & 0.500 & 16QAM & 0.500 & 16 & 173 & 28.82 \\ \hline
8  & QPSK & 0.500 & 16QAM & 0.500 & 16 & 173 & 28.82 \\ \hline

\end{tabular}
\end{minipage}

&
\begin{minipage}{0.46\linewidth}
\centering
\textbf{(b) H.265 (Baseline)} \\[4pt]

\begin{tabular}{|c|c|c|c|c|c|}
\hline
\textbf{SNR} 
& \textbf{Mod}
& \textbf{Rate}
& \textbf{Source Bitrate}
& \textbf{PSNR} \\
\hline

-2 & QPSK   & 0.500 & 165 & Failed   \\ \hline
0  & QPSK   & 0.500 & 165 & Failed   \\ \hline
2  & QPSK   & 0.500 & 165 & Failed   \\ \hline
4  & QPSK   & 0.500 & 173 & Failed   \\ \hline
6  & 16QAM  & 0.500 & 173 & 11.24 \\ \hline
8  & 16QAM  & 0.500 & 173 & 27.77 \\ \hline

\end{tabular}
\end{minipage}

\end{tabular}

\end{table*}

We further compare the proposed semantic source--channel coding framework with an adaptive H.265 benchmark on both the MCL-JCV and UVG datasets. For a fair comparison, the H.265 encoder is configured to match the overall bitrate and resource budget of our method, where the MCL-JCV experiments operate under a 350\,kHz bandwidth and the UVG experiments under a 1060\,kHz bandwidth. Specifically, we employed the x265 (libx265) software implementation~\cite{x265} of H.265 via FFmpeg as the baseline video codec. The encoder was operated in constant bitrate mode, with the bitrate configured to match that of the corresponding TokenCom configuration for each (video, SNR) pair. As summarized in Table~\ref{tab:parameter_compare}, the H.265 baseline employs conventional pixel-domain source coding with a single MCS per frame, whereas our TokenCom system performs content-aware token coding with semantic-driven bit allocation and per-class UEP under an identical resource budget. As shown in Fig.~\ref{fig:curve_h265}(a)(b), the proposed method consistently outperforms adaptive H.265 in all perceptual and semantic metrics, including LPIPS, CLIP similarity, and FVD, across every SNR level and for both resolutions. The semantic and perceptual quality gains are substantial: our token-based scheme achieves significantly lower LPIPS, higher CLIP similarity, and markedly lower FVD, indicating superior semantic fidelity and temporal coherence. According to the adopted evaluation protocol, an operating point is considered valid only if at least 85\% of the frames can be successfully reconstructed. This threshold is introduced as a pragmatic design choice to ensure a minimum level of temporal continuity and perceptual usability in video playback, and is consistently applied to all compared schemes. At severely low SNRs, the adaptive H.265 pipeline often experiences severe packet losses, leading to unsuccessful decoding for a large fraction of frames, thereby in Fig. \ref{fig:curve_h265}, the severely low SNR points had to be excluded from the H.265 curve. In contrast, the proposed framework maintains stable decodability across all SNRs and both bandwidths. These results demonstrate that the proposed semantic source--channel coding framework offers substantially more robust and perceptually faithful video reconstruction than H.265, particularly when evaluated under perceptual or semantic criteria and across both medium and high resolution video settings. Moreover, the per-SNR selections are provided in Table~\ref{tab:optimized_compare} under 340 kHz BW, confirm that the proposed optimizer effectively adapts the bit-precision, modulation, and coding rate in response to channel quality. Under different bandwidth budgets, the algorithm consistently identifies MCS configurations that fully utilize the available resources while minimizing distortion, demonstrating the practical effectiveness of the source--channel optimization. Finally, at the SNR of 8~dB, the proposed video TokenCom framework achieves a channel bandwidth ratio (CBR) of $2.2\times10^{-3}$ for the video sequence\textit{``The woman is hitting the man's mobile phone''}, evaluated on the first 16 frames under a bandwidth of 340~kHz.

\begin{figure*}[t]
    \centering
    \includegraphics[width=0.95\textwidth]{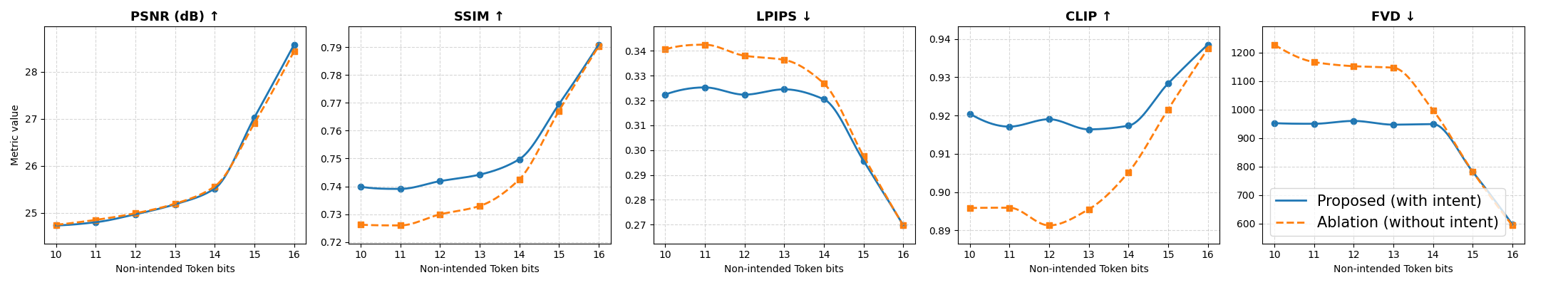} 
    \caption{The ablation results for the proposed video TokenCom framework with and without textual intent guidance on the MCL-JCV dataset.}
    \label{fig:ablation} 
\end{figure*}

\subsection{Computational delay and overhead}
To evaluate the computational delay of our method, we conducted experiments on a single NVIDIA A6000 GPU using a 1000-frame video at a resolution of 1024$\times$640 as input. For full-bit-precision processing, the average computation time per frame was 65\,ms, while for other bit-precision the per-frame time was 122\,ms. In addition, we measured the computational complexity of the model. For a 1024$\times$640 input video, the model requires $5{,}424\ \text{GFLOPs}$ per frame and contains $247.806\ \text{M}$ learnable parameters.

The communication overhead of transmitting the semantic mask to the receiver is negligible compared to the video payload. For example, for a 1024$\times$640 video segment with 17 frames, the overhead of explicitly signaling the semantic mask is calculated according to equation~\eqref{eq:overhead}, with $B_{\mathrm{full}}=16$, $B_\Delta=11$, block size of 4$\times$4, and an intended token ratio $\rho_s=0.7$, the mask represents approximately $1.7\%$ of the communication payload. Moreover, this overhead can be further reduced through simple run-length or block-based coding. 

\subsection{Ablation Study}
As an ablation study, to demonstrate effectiveness of our proposed textual intent-aware rate adaptation, we compared with an intent-unaware ablation which removes the CLIP and optical flow modules and selects the non-intended tokens randomly instead. The quantitative results under the MCL-JCV dataset are presented in Fig.~\ref{fig:ablation}. We tested seven different $B_{\Delta}$ values, namely $10$, $11$, $12$, $13$, $14$, $15$, and $16$. As expected, by increasing $B_{\Delta}$, all metrics exhibit an overall improving trend, with similar performance for our proposed scheme (with intent) and the intent-unaware ablation (without intent) at full precision. However, at limited rate values with low $B_{\Delta}$, the proposed intent-aware scheme achieves clear benefits in semantic fidelity and perceptual quality as demonstrated in CLIP, LIPIPs, and FVD metrics. Note that these three metrics evaluate intent-related semantic preservation, while the PSNR and SSIM merely evaluate pixel-level distortions. Thereby, the improvements in CLIP, LIPIPs, and FVD metrics demonstrate the semantic preservation gains by our proposed textual intent-awareness scheme under constrained bitrate.

\section{Conclusion}
This paper presented a multi-rate video token communication framework driven by textual intent, enabling semantic- and content-aware rate adaptation for efficient multimedia transmission in future AI-native wireless networks. By integrating text-conditioned heatmaps, optical-flow propagation, and video tokenization, we established a fine-grained mapping between user-intended semantics and discrete video tokens. A multi-rate bit-precision allocation strategy was developed, allocating high precision indices from a token codebook to user-intended tokens while employing low-bit differential indexing from a reduced codebook for non-intended tokens to reduce the transmission overhead. Furthermore, a source--channel adaptation scheme under unequal error protection was designed to balance distortion and delay within a fixed resource budget. Extensive experiments on various video datasets demonstrate that the proposed video TokenCom framework outperforms both conventional and semantic communication baselines in perceptual and semantic quality metrics on a wide range of SNRs. Notably, compared with conventional H.265 video compression and transmission scheme, our proposed Video TokenCom achieves consistent improvements in PSNR, SSIM, LPIPS, FVD, and CLIP-based semantic similarity. In particular, at a typical channel SNR value of 6 dB, Video TokenCom reduces the FVD metric by nearly 1500.


\ifCLASSOPTIONcaptionsoff
  \newpage
\fi

\bibliographystyle{IEEEtran}
\bibliography{refs}

\end{document}